\documentclass[
showpacs,preprintnumbers,amsmath,amssymb]{revtex4}
\usepackage{amsmath}
\usepackage{graphicx}
\begin{document}
\title{Reconstructing cosmological solutions in F(R) gravity. Towards a unified model of the Universe evolution}
\author{Diego S\'aez-G\'omez$^1$}
\address{$^1$Institut de Ci\`{e}ncies de l'Espai
ICE/CSIC-IEEC, Campus UAB, Facultat de Ci\`encies, Torre
C5-Parell-2a pl, E-08193 Bellaterra (Barcelona) Spain}
\begin{abstract}
It is accepted by the majority of scientific community that the Universe is currently in an accelerated epoch. In order to explain this {\it shock} of late 90's, a lot of dark energy candidates have been proposed. We study in the context of $f(R)$ gravity, how some modifications on General Relativity could reproduce such behavior of the cosmic evolution. It is showed that in general, an $f(R)$ theory can be reconstructed from a non-minimal scalar-tensor theory, where a desired cosmological solution can be achieved. Some viable models are studied as well as its cosmological evolution, and the possibility that inflationary epoch is also a result of these fine modifications.
\end{abstract}
\maketitle
\section{Reconstructing  F(R) gravity from scalar-tensor theory}
In this section, it is  showed how in general an arbitrary cosmological solution can be reconstructed in f(R) gravity from  a non-minimal coupled scalar-tensor theory (see Ref.\cite{<FRandST1>} and \cite{<FRandST2>}). Let us start by writting the action that describes a general non-minimal scalar-tensor theory,
\begin{equation}
S=\int d^4x\sqrt{-g}P(\phi)R+Q(\phi)+L_m\ ,
\label{1.1}
\end{equation}
where $L_m$ is the lagrangian for the matter fields. Note that the scalar field $\phi$ has no kinetic term for convenience in order to be able to reconstruct $f(R)$ gravity from the action (\ref{1.1}). By  varying the action on the metric tensor $g_{\mu\nu}$ and on the scalar field $\phi$, the gravity and scalar field equations  are obtained,
\[
\frac{1}{2}g_{\mu\nu}\left( P(\phi)R+Q(\phi)\right)+P(\phi)R_{\mu\nu}+g_{\mu\nu}\nabla_{\sigma}\nabla^{\sigma} P(\phi)-\nabla_{\mu}\nabla_{\nu}P(\phi)= \frac{\kappa^2}{2}T^{(m)}_{\mu\nu}\ ,
\]
\begin{equation}
P'(\phi)R+Q'(\phi)=0\ ,
\label{1.2}
\end{equation}
here the primes denote derivatives respect $\phi$. The second equation may be resolved  with the scalar field as a function of $R$, $\phi=\phi(R)$, and by replacing this result in the action (\ref{1.1}), the action for $f(R)$ is obtained with $f(R)=P\left(\phi(R)\right)R+Q\left(\phi(R)\right)$, and it yields $S=\int d^4x\sqrt{-g}\left(f(R)+L_m\right)$, which is the general action for $f(R)$ gravity, and whose field equation are obtained by varying the action respect the metric $g_{\mu\nu}$,  
\begin{equation}
R_{\mu\nu}f'(R)-\frac{1}{2}g_{\mu\nu}f(R)+g_{\mu\nu}\nabla_{\sigma}\nabla^{\sigma} f'(R)-\nabla_{\mu}\nabla_{\nu}f'(R)=\frac{\kappa^2}{2}T^{(m)}_{\mu\nu}\ ,
\label{1.2a}
\end{equation}
Hence, in general, any cosmological model could be solved by the first field equation in (\ref{1.2}) and the function $f(R)$ is reconstructed by the scalar field equation. We are interested in solutions for a flat FRW Universe, this means for a metric $ds^2=-dt^2+a^2(t)\sum^{3}_{i=1}dx_{i}^2$, in such a case the Friedmann equations for (\ref{1.2}) read as \cite{<FRandST2>},
\[
3H\frac{dP(\phi)}{dt}+3H^2P(\phi)+\frac{1}{2}Q(\phi)-\frac{\rho_m}{2}=0\ , 
\]
\begin{equation}
2\frac{d^2P(\phi)}{dt^2}+4H\frac{dP(\phi)}{dt}+(4\dot{H}+6H^2)P(\phi)+Q(\phi)+p_m=0\ .
\label{1.4}
\end{equation}
In order to get a general solution for the Hubble parameter , $H(t)=\dot{a}/a$, we redefine for simplicity the scalar field  to be the time coordinate $\phi=t$. The perfect fluid defined by $T^{(m)}_{\mu\nu}$ has an equation of state (EoS) given by $p_m=w_m\rho_m$,  and its evolution is described by the continuity equation $\dot{\rho}_m +3H(1+w_m)\rho_m=0$. Then, the Hubble parameter may be calculated as a function of the scalar field $\phi$, $H=g(\phi)$, and by combining equations (\ref{1.4}), the following solution is obtained
\begin{equation}
g(\phi)=-\sqrt{P(\phi)}\int d\phi \frac{P''(\phi)}{2P^{3/2}(\phi)}+ H_0\sqrt{P(\phi)}\ ,
\label{1.8}
\end{equation}
where $H_0$ is an integration constant, and we have considered no matter contribution for simplicity. Then, the solution for the Hubble parameter is calculated $H=g(\phi(t))=g(t)$. As we have fixed the solution for the scalar field to be $\phi=t$, the  equations in (\ref{1.4}) act as a constraint on $Q(\phi)$. Let us consider a simple example to show how one could find a cosmological solution in scalar-tensor theory, and the corresponded $f(R)$ function. We consider (see Ref.\cite{<FRandST1>}) $P(\phi)=\phi^2$, the solution is given by (\ref{1.8}),
\begin{equation}
H(t)=H_0t+\frac{1}{2}\frac{1}{t}\ ,
\label{1.9}
\end{equation}
which describes a Universe that passes through two phases, a first decelerated epoch identified as the radiation/matter dominated epochs, and a second accelerated, which corresponds with the current epoch.  
The function $Q(\phi)$ is defined by (\ref{1.7}), and by inverting the scalar field equation (\ref{1.2}), it is calculated $\phi=\phi(R)$, and the function  $f(R)$ yields,
\begin{equation}
f(R)=\left[R-6(H_0(H_0+1)+5/2) \right]\frac{R-2k\pm\sqrt{R(1-2H_0)}}{2}+const\ .
\label{1.20}
\end{equation}
Thus, this expression for the function $f(R)$ reproduces the current cosmic acceleration with no need to introduce any kind of dark energy. Nevertheless, the modification of the law of gravity implies other serious problems as the violation of local tests, which can be avoided for a class of $f(R)$ theories called viable $f(R)$ gravity that are studied in the next section.   

\section{Cosmological evolution from viable f(R) gravity}
The modifications introduced by the function $f(R)$ in General Relativity may resolve the dark energy problem, although it could  imply other problems as the violation of Solar System tests, where GR is very well constrained, or problems on stability around matter distributions. Some models have been proposed where these problems are avoided (see Ref.\cite{FRViable} and \cite{FRViable2}), and where also the inflation can be achieved. This kind of models assumes a theory of gravity described by an action $S=\int d^4\sqrt{-g}R+ F(R)$, such that $F(R)$ is negligible for small scales (and the Hilbert-Einstein action is recovered) but it starts to become important for large scales, where its  effects contributes to the cosmic acceleration. Let us consider this kind of action with $F(R)$ given by 
\begin{equation}
F(R)= \frac{R^2(\alpha R^2-\beta)}{1+\gamma R^2}\ .
\label{2.1}
\end{equation}
This model has been analyzed in Ref.\cite{DSG}, where it is found that passes through at least two de Sitter points, which can be corresponded with the inflation and cosmic acceleration eras (the idea of unified inflation and cosmic acceleration under  $f(R)$ theory was suggested in Ref.\cite{InfCosm}). Let us write the first Friedmann equation for a flat FRW universe, from (\ref{1.2a}) with $f(R)= R+ F(R)$, it yields for the function (\ref{2.1})
\[3H^2=-\frac{R^2(\alpha R^2-\beta)}{2(1+\gamma R^2)}+3(H^2+\dot{H})\frac{2R(\alpha\gamma R^4-2\alpha R^2-\beta)}{(1+\gamma R^2)^2}-
 \]
\begin{equation}
-18F''(R)(H^2\dot{H}+H\ddot{H})\frac{2\alpha\gamma^2 R^6+20\alpha\gamma R^4+6(\beta\gamma-\alpha)R^2-2\beta}{(1+\gamma R^2)^3}\ ,
\label{2.2}
\end{equation}
where the matter has been neglected for simplicity. This equation can be rewritten as a dynamical system with $\dot{H}=C$, $\dot{C}=F_1(H,C)$, where the critical points are clearly those for $\dot{H}=0$, i.e. de Sitter points (see Ref.\cite{DynamSys}). For our model, the existence of these critical points can be achieved by studying the equation (\ref{2.2}) for $H=H_0$ a constant, in such a case the equation reduces to
\begin{equation}
\frac{\gamma}{4}R_0^5-\gamma\beta R_0^4+\frac{\gamma}{2}R_0^3+\frac{1}{4}R_0=0\ ,
\label{2.3}
\end{equation}
where $R_0=12H^2_0$. The positive roots for $R_0$ correspond to de Sitter points. By a simple analysis of equation (\ref{2.3}), and using the Descartes' rule of signs, one concludes that the equation (\ref{2.3}) can have either two or no positive roots, where each one can be identified with the inflation and cosmic acceleration epoch. As the function (\ref{2.1}) posses a minimum, which can be seen as an effective cosmological constant, it is clear that Universe passes through two de Sitter points. Let us now study in more detail this model during the current epoch, when matter is included $p_m=w_m\rho_m$. As it was pointed above, the function (\ref{2.1}) has a minimum, which can be easily calculated assuming $\beta\gamma/\alpha>>1$. It gives $R_0 \sim( \beta/\alpha\gamma)^{1/4}$ where $F'(R_0)=0$ and  $F(R_0)= -2\tilde{R_0}\sim -\frac{\beta}{\gamma}$. Then, we can study the cosmological evolution around this point $R=R_0$, where the function (\ref{2.1}) can be expressed as $F(R)=-2\tilde{R_0}+f_0(R-R_0)^2+O\left( (R-R_0)^3\right)$. The solution for the Friedmann equation will have the form $H(t)=H_0(t)+\delta H$, which at zero order is the same as in the case of a cosmological constant,
\begin{equation}
H^2(z)=\frac{\tilde{R_0}}{3}\left[ A^2(1+z)^{3(w_m+1)}+1\right]\ .
\label{2.5}
\end{equation}
where the Hubble parameter is expressed as a function of the redshift $z$ instead of time. The perturbations $\delta H$ are irrelevant around the minimum $R_0$. This solution is the same to $\Lambda$CDM model, although the evolution of the cosmological parameters is not exactly the same, as it is showed in Fig.1. We can fit the free parameters of the model ($\alpha,\beta,\gamma$) by using the observable values ($H_0$, $\Omega^0_m$) at the current epoch. Then, by taking the minimum $R_0$ at $z=0$, we can study the cosmological evolution for small redshifts, where the derivatives of $F(R)$ are negligible compared with the value of the function and the solution (\ref{2.5}) is valid. In order to account the effects of the geometrical terms from $F(R)$, an effective cosmological parameter is defined,
\begin{equation}
\Omega_{F(R)}(z)=\frac{\rho_{F(R)}}{3H^2(z)}=-\frac{F(R)}{6H^2(z)}+\left[ 1+\frac{\dot{H}(z)}{H^2(z)}\right] F'(R)-18F''(R)\left[ \dot{H}(z)+\frac{\ddot{H}(z)}{H(z)}\right]\ ,
\label{2.6}
\end{equation}
This function is plotted in Fig.1 versus the redshift, and compared with $\Lambda$CDM model for $z=0$ to $z=1.8$. Note that the evolution is very similar, both models have common points at $z=0$ and $z=1.74$, while the behavior between such points is completely different. This difference comes form the fact that while a cosmological constant posses a constant EoS parameter, in $F(R)$ gravity, the EoS parameter defined as $w_{F(R)}=p_{F(R)}/\rho_{F(R)}$, changes with the redshift, as it goes from -1 for $z=0$ to 0 for $z=1.5$.
\begin{figure}[h]
 \centering
 \includegraphics[width=3in,height=2in]{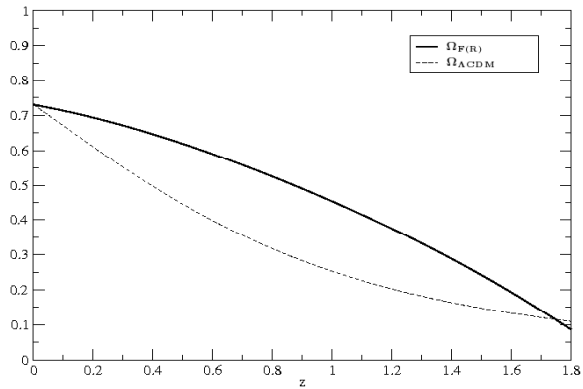}
 \caption{Evolution of the cosmological parameter from dark energy versus redshift, such for $F(R)$ theory as for $\Lambda CDM$ model.}
 \label{fig1}
\end{figure}
\section{Conclusions}
It has been showed that a general solution in a flat FRW Universe can be achieved in $f(R)$ gravity with the help of an auxiliary scalar field. Then, the acceleration epochs of the Universe could be explained just in the context of gravity with no need to introduce new forms of the so-called dark energy. As the modifications on GR produce some undesirable effects, one has to consider  some complex  $f(R)$ functions which could avoid them. In this context, the current cosmological evolution in $f(R)$ is compared with the popular $\Lambda$CDM model, and it is seen that a similar behavior is achieved for low redshifts. Alternative methods of reconstruction have been recently proposed in Ref.\cite{FRrecons}, which could imply a more powerful technique and give interesting properties.

\end{document}